\documentclass[superscriptaddress,aps,preprintnumbers,amsmath,showpacs,amssymb,prd,nofootinbib,reprint]{revtex4-1}
\usepackage{amsfonts}
\usepackage[mathscr]{eucal}
\usepackage{ascmac}
\usepackage{dcolumn}
\usepackage{bm}
\usepackage[colorlinks=true,linkcolor=blue,citecolor=blue]{hyperref}
\usepackage{hyperref}
\usepackage{color}

\begin{document}


\def\a{\alpha}
\def\b{\beta}
\def\c{\varepsilon}
\def\d{\delta}
\def\f{\phi}
\def\g{\gamma}
\def\h{\theta}
\def\k{\kappa}
\def\l{\lambda}
\def\m{\mu}
\def\n{\nu}
\def\p{\psi}
\def\q{\partial}
\def\r{\rho}
\def\s{\varphi}
\def\t{\tau}
\def\u{\upsilon}
\def\v{\varphi}
\def\w{\omega}
\def\x{\xi}
\def\y{\eta}
\def\z{\zeta}
\def\D{\Delta}
\def\G{\Gamma}
\def\H{\Theta}
\def\L{\zeta}
\def\F{\Phi}
\def\P{\Psi}
\def\S{\Sigma}

\def\aa{{\dot \a}}
\def\bb{{\dot \b}}
\def\ss{{\bar \s}}
\def\hh{{\bar \h}}
\def\CA{{\cal A}}
\def\CB{{\cal B}}
\def\CC{{\cal C}}
\def\CD{{\cal D}}
\def\CE{{\cal E}}
\def\CG{{\cal G}}
\def\CH{{\cal H}}
\def\CI{{\cal I}}
\def\CK{{\cal K}}
\def\CL{{\cal L}}
\def\CR{{\cal R}}
\def\CM{{\cal M}}
\def\CN{{\cal N}}
\def\CO{{\cal O}}
\def\CP{{\cal P}}
\def\CQ{{\cal Q}}
\def\CS{{\cal S}}
\def\CT{{\cal T}}
\def\CW{{\cal W}}

\newcommand{\Slash}[1]{{\ooalign{\hfil/\hfil\crcr$#1$}}}

\def\o{\over}
\newcommand{\gsim}{ \mathop{}_{\textstyle \sim}^{\textstyle >} }
\newcommand{\lsim}{ \mathop{}_{\textstyle \sim}^{\textstyle <} }
\newcommand{\vev}[1]{ \left\langle {#1} \right\rangle }
\newcommand{\bra}[1]{ \langle {#1} | }
\newcommand{\ket}[1]{ | {#1} \rangle }
\newcommand{\EV}{ {\rm eV} }
\newcommand{\KEV}{ {\rm keV} }
\newcommand{\MEV}{ {\rm MeV} }
\newcommand{\GEV}{ {\rm GeV} }
\newcommand{\TEV}{ {\rm TeV} }
\def\diag{\mathop{\rm diag}\nolimits}
\def\Spin{\mathop{\rm Spin}}
\def\SO{\mathop{\rm SO}}
\def\O{\mathop{\rm O}}
\def\SU{\mathop{\rm SU}}
\def\U{\mathrm{U}}
\def\Sp{\mathop{\rm Sp}}
\def\USp{\mathop{\rm USp}}
\def\SL{\mathop{\rm SL}}
\def\tr{\mathop{\rm tr}}
\def\rank{\mathop{\rm rank}}

\def\spin{\mathfrak{ spin}}
\def\so{\mathfrak{so}}
\def\o{\mathfrak{o}}
\def\su{\mathfrak{su}}
\def\u{\mathfrak{u}}
\def\sp{\mathfrak{sp}}
\def\usp{\mathfrak{usp}}
\def\sl{\mathfrak{sl}}
\def\e{\mathfrak{e}}

\def\beq#1\eeq{\begin{align}#1\end{align}}
\def\alert#1{{\color{red}[#1]}}

\renewcommand{\baselinestretch}{0.97}


\preprint{
IPMU 16-0006; 
}

\title{
A search for minimal 4d $\mathcal{N}=1$ SCFT 
}

\author{
Dan Xie
}
\affiliation{Center of Mathematical Sciences and Applications, Harvard University, Cambridge, 02138, USA}
\affiliation{Jefferson Physical Laboratory, Harvard University, Cambridge, MA 02138, USA}

\author{
Kazuya Yonekura
}
\affiliation{Kavli IPMU (WPI), UTIAS, 
The University of Tokyo, 
Kashiwa, Chiba 277-8583, Japan}


\begin{abstract} 
We discuss a candidate for a minimal interacting 4-dimensional $\mathcal{N}=1$ superconformal field theory (SCFT).
The model contains a chiral primary operator $u$ satisfying the chiral ring relation $u^2=0$, and its scaling dimension is $\Delta(u)=1.5$.
The model is derived by turning on a $\CN=1$ preserving deformation of $\CN=2$ $A_2$ Argyres-Douglas theory.
The central charges are given by $(a,c)=(263/768, 271/768) \simeq (0.342,0.353)$.
There is no moduli space of vacua, no flavor symmetry, and the chiral ring is finite.

\end{abstract}

\maketitle


\section{Introduction
\label{sec:introduction}}
Recent progress of the conformal bootstrap program in dimensions larger than 2 (initiated in \cite{Rattazzi:2008pe}) has provided many invaluable insights to the study of 
quantum field theory. 

One of the interesting directions in those studies is to find a minimal interacting conformal field theory (CFT) in each dimension.\footnote{
In this letter we do not specify the precise measure of minimality of a theory. 
Bootstrap studies often constrain the central charge $c$ defined by the two point function of the energy-momentum tensor.
But it is natural to use $a$ in 4d or $F$ in 3d as a measure of the degrees of freedom, from the point of view of the $a$-theorem~\cite{Komargodski:2011vj} 
and $F$-theorem~\cite{Jafferis:2011zi,Casini:2012ei}.
In fact, the $c$ can increase along renormalization 
group flows in 4d~\cite{Cappelli:1990yc,Anselmi:1997am} and in 3d \cite{Nishioka:2013gza}.}
Candidates for such a minimal CFT might be Ising model in $(d=3,\CN=0)$~\cite{ElShowk:2012ht}, 
Wess-Zumino model in $(d=3,\CN=2)$~\cite{Bobev:2015vsa}, 
$A_2$ Argyres-Douglas theory in $(d=4,\CN=2)$~\cite{Liendo:2015ofa}, $\SU(2)$ gauge theory in $(d=4,\CN=4)$~\cite{Beem:2013qxa}, 
$A_1$ theory in $(d=6, \CN=(2,0))$~\cite{Beem:2015aoa}, and so on.

In the cases listed above, it is not difficult to guess the minimal theory from the theories known in the literature.
However, in the $(d=4, \CN=1)$ case, even a possible candidate for such a minimal theory has not been so clear.
In this letter, we initiate a search for such a theory. 

Bootstrap searches of a minimal 4d SCFT~\cite{Poland:2011ey,Bobev:2015jxa,Poland:2015mta} suggest that 
such a theory might have a chiral operator $u$ satisfying the chiral ring relation $u^2=0$. 
The main purpose of this letter is to construct an explicit model having this chiral ring relation.
The model also has rather small central charges $a$ and $c$; it is smaller than those of any known Lagrangian fixed points such as supersymmetric (SUSY) QCD.
We hope that our work motivates a study of either (i) finding a smaller theory than the model presented in this letter, or (ii) making the bootstrap constraints sharper
so as to confirm the minimality of the model.

\section{Field theory construction}
One of the properties~\cite{Poland:2011ey,Bobev:2015jxa,Poland:2015mta} which might be satisfied by a putative minimal SCFT 
is that it contains a chiral primary operator $u$ satisfying 
the chiral ring relation $u^2=0$. Our purpose is to construct a model satisfying that chiral ring relation.
In such a model, the scaling dimension of $u$, which we denote as $\Delta(u)$, is currently bounded as 
$\Delta(u)> 1.41....$~\cite{Poland:2011ey,Bobev:2015jxa,Poland:2015mta}. 
The bound on $\Delta(u)$ is obtained by numerical
analysis and may become sharper in the future analysis of conformal bootstrap. 
It turns out that $\Delta(u)=1.5$ in our model.
The model is obtained by a simple deformation of $A_2$ Argyres-Douglas (AD) theory~\cite{Argyres:1995jj}.\footnote{ Some deformation of $\mathcal{N}=2$ AD theory was already  considered in \cite{Argyres:1995jj} (see also \cite{Eguchi:2003wv}).  
They considered the induced deformation on AD theory by turning on a $\mathcal{N}=1$ deformation of UV $\SU(3)$ gauge theory by $\tr \Phi^3$. 
Their deformation is different from ours.}

Let us first consider the $\CN=2$ AD theory~\cite{Argyres:1995jj,Argyres:1995xn}.
This theory contains an $\CN=2$ chiral supermultiplet, and in terms of $\CN=1$ supersymmetry, it contains two scalar chiral primaries and one spinor chiral primary
as $(u, \lambda_\alpha, S)$. Here $u$ is the usual $\mathcal{N}=2$ Coulomb moduli field.

$\CN=2$ SCFTs have R-symmetries $\U(1)_R \times \SU(2)_R$. Let us denote the Cartan generators of these R-symmetries as $R_{\U(1)}$ and $R_{\SU(2)}$,
respectively. The R-charges of $u$ are determined in \cite{Argyres:1995xn} and the R-charges of $\lambda_\alpha$ and $S$ follow from $\CN=2$ supersymmetry.
They are listed in table~\ref{tab:Rc}.
In particular, the scaling dimension of $u$ at  $\CN=2$ superconformal fixed point is given by 
\beq
\Delta_{\CN=2}(u)=\frac{1}{2}R_{\U(1)}(u)=\frac{6}{5}.
\eeq


Now we consider $\CN=1$ deformation of this theory. We introduce a superpotential of the form
\beq
W \sim u^2. \label{eq:deform}
\eeq
From the above scaling dimension of $u$ in the $\CN=2$ theory, we can see that this is a relevant deformation.
We give some evidence that the theory flows to another nontrivial superconformal fixed point in the infrared (IR).

Assuming that there is no accidental $\U(1)$ global symmetry in the IR, the $\CN=1$ R-charge is given by a linear combination as $x  R_{\U(1)}+(1-x)  R_{\SU(2)}$
for some parameter $x$.
The superpotential \eqref{eq:deform} preserves only a particular combination of the R-symmetries given by
\beq
R=\frac{5}{12} R_{\U(1)} +\frac{7}{12} R_{\SU(2)}.\label{eq:Rc}
\eeq
The R-charges of chiral operators under this $R$ are also listed in table~\ref{tab:Rc}. 
Assuming that the theory has a superconformal fixed point without accidental $\U(1)$ symmetries, the scaling dimensions of chiral primary operators
are related to the R-charges as 
\beq
\Delta=\frac{3}{2}R.
\eeq
One can see that all the chiral operators listed in table~\ref{tab:Rc} satisfy the usual unitarity bounds $\Delta(u), \Delta(S) > 1$ (for scalar operators)
and $\Delta({\lambda_\alpha})>3/2$ (for spinor operators).
This is one of the supports of our claim that the theory has a nontrivial fixed point.
At the fixed point, we get the scaling dimension of $u$ as
\beq
\Delta(u)=\frac{3}{2}=1.5
\eeq

\begin{table}[t]
\centering
\begin{tabular}{|c|c|c|c|}
\hline
& $u$ & $\lambda_\alpha$ & $S$ \\ \hline
$R_{\U(1)}$ & $12/5$ & $7/5$ & $2/5$  \\ \hline
$R_{\SU(2)}$ & $0$ & $1$ & $2$ \\ \hline
$R$ & $1$ & $7/6$ & $4/3$ \\ \hline
\end{tabular}
\caption{R-charges of chiral operators.}
\label{tab:Rc}
\end{table}

Let's now argue that our model is the minimal interacting SCFT among the theories which can be obtained from deformation of $A_2$ AD theory. The $A_2$ AD theory has 
three relevant scalar operators $(u, S, u^2)$ with increasing scaling dimensions. The $\mathcal{N}=1$
preserving deformation by $u$ gives us a free chiral field in the IR \cite{Bolognesi:2015wta}.
Actually, one can check that the operator $S$ hits the unitarity bound in this case. 
On the other hand, the deformation by $S$ is the usual $\mathcal{N}=2$ preserving deformation and one gets a free $\mathcal{N}=2$ vector multiplet in the IR. 
Therefore, only the $u^2$ deformation gives us a nontrivial SCFT.  

One might wonder  if we could get more minimal theory by starting with other $\mathcal{N}=2$ AD theories and 
turning on $\mathcal{N}=1$ preserving relevant deformations of the form $W \sim u^2$ with $u$ a Coulomb branch operator. 
An easy computation shows that our model has the minimal central charge 
among those possibilities. We will comment on the possibility of deformation $W \sim u$ in more general AD theories later.

\subsection{Chiral ring relation}
Now we show the chiral ring relation $u^2=0$ in the above model.
What we will actually show is that $u^2$ is trivial as an element of the chiral ring,
\beq
u^2= \bar{D}^2(\cdots), \label{eq:trivial}
\eeq
where the superspace notation follows that of Wess and Bagger.
This equation holds once we introduce the superpotential \eqref{eq:deform}, even without going to the IR fixed point.
Then, at the IR fixed point, 
the usual argument in the superconformal field theory tells us that $u^2(x):=\lim_{y \to x} u(x) \times u(y)=0$ in the operator product expansion (OPE) of $u \times u$.

To show \eqref{eq:trivial}, let us consider the following more general situation. Suppose that we have an $\CN=1$ SCFT $\CS_{\rm UV}$
with a $\U(1)$ flavor symmetry which we denote as $\U(1)_F$. Let $J$ be the current supermultiplet of this flavor symmetry.
This operator satisfies the conservation equation
\beq
\bar{D}^2 J=0.\label{eq:conserve}
\eeq

Now we deform this theory by adding a superpotential
\beq
W=\eta \CO
\eeq
where $\CO$ is some chiral superfield and $\eta$ is a parameter. If $\eta$ is small, we can treat it as a perturbation. 
Then, in the deformed theory, the current conservation equation \eqref{eq:conserve} is modified as
\beq
\bar{D}^2 J=\delta_F W=q_F \eta \CO,
\eeq
where $\delta_F$ means the transformation under $\U(1)_F$, and $q_F$ is the $\U(1)_F$ charge of $\CO$. 
This equation shows that $\CO$ is a trivial element of chiral ring of the theory once we turn on nonzero $\eta$ if $q_F \neq 0$.
Although this equation is derived by assuming that $\eta$ is small, the qualitative feature that two protected operators $J$ and $\CO$
combine to become an unprotected operator should be true for any finite $\eta$.

We can apply the above general argument to our theory. The UV theory $\CS_{\rm UV}$ is the $\CN=2$ AD theory.
We take the $\U(1)_F$ charge $F$ as $F=R_{\U(1)}- R_{\SU(2)}$ which is a flavor symmetry in terms of $\CN=1$ supersymmetry.
We also take the operator $\CO$ as $\CO=u^2$. This establishes the relation \eqref{eq:trivial}.
Note that there is no flavor symmetry in the IR because $F=R_{\U(1)}- R_{\SU(2)}$ is broken.

Let us remark the following crucial point. Suppose that we have a chiral field $\Phi$ which is an elementary field in a Lagrangian field theory.
If we introduce a superpotential $W \sim \Phi^2$, then not only $\Phi^2$ but also $\Phi$ itself becomes trivial because we can simply
integrate out the field $\Phi$. Even if $\Phi$ is a composite field, it often happens that there exists Seiberg duality 
such that $\Phi$ becomes an elementary field in the dual side when the dimension $\Delta(\Phi)$ is rather small.
In our construction, it is crucial that the AD theory is a kind of ``non-Lagrangian'' theory in the sense that $u$ does not allow
any interpretation as an elementary chiral field, but nevertherless $u$ has a small scaling dimension
($\Delta^{\CN=2}(u)=6/5$) such that $W \sim u^2$ is a relevant deformation.
Note that the $u$ must be nontrivial because it can trigger an RG flow to a free theory~\cite{Bolognesi:2015wta}.

\subsection{Minimal models in 4d SCFT}
We would like to point out an interesting feature of our model;
it only has finitely many chiral primary operators. 

First let us consider the chiral ring of the $\CN=2$ $A_2$ AD theory which is found at 
a point in the Coulomb branch of $\SU(2)$ gauge theory with $N_f=1$ flavor~\cite{Argyres:1995xn}. 
The chiral operators in the $\SU(2)$ theory are given as
\beq
u \sim \tr \Phi^2, ~~ \lambda_\alpha \sim \tr ( \Phi W_\alpha), ~~ S \sim   \tr (W^\alpha W_\alpha) ,
\eeq 
where $\Phi$ is the adjoint chiral field and $W_\alpha$ is the field strength chiral multiplet.
At the classical level (see e.g., \cite{Cachazo:2002ry}), a chiral operator $q$ in a representation $\rho$ of a gauge group $G$ satisfies $\rho(W_\alpha) q = 0$ as a chiral ring,
and in particular we have $W_\alpha \Phi-\Phi W_\alpha = 0$ and $W_\alpha W_\beta+W_\beta W_\alpha = 0$.
Furthermore, $\su(2)$ matrices $A, B, C, D$ satisfy $2 \tr (ABCD)=\tr(AB)\tr(CD)-\tr(AC)\tr(BD)+\tr(AD)\tr(BC)$.
Using these relations, we get classical ring relations
\beq
\lambda^\alpha \lambda_\alpha=uS,~~\lambda_\alpha S=0,~~S^2=0. \label{eq:CCR}
\eeq
In general, chiral ring relations can be modified quantum mechanically. However, it is reasonable to assume that there is a quantum chiral ring relation corresponding to each
classical chiral ring relation. From the fact that quantum corrections must respect $\U(1)_R \times \SU(2)_R$ R-charges listed in Table~\ref{tab:Rc}, 
we conjecture that \eqref{eq:CCR} is valid quantum mechanically at the AD fixed point, up to corrections to the coefficient of $\lambda^\alpha \lambda_\alpha=uS$.
One can also check that chiral operators involving the quarks of the $\SU(2)$ $N_f=1$ theory can be replaced by $u, \lambda_\alpha$ and $S$
by using Konishi anomaly equations. For example, the $\CN=2$ current multiplet of $\U(1)$ flavor symmetry is actually a
descendant as pointed out in \cite{Argyres:1995xn}, and in $\CN=1$ language one has $\tilde{Q}Q = S$, where $Q$ and $\tilde{Q}$ are quarks.

After turning on the deformation \eqref{eq:deform}, the operator $u$ satisfies $u^2=0$ as shown above. 
Therefore, the only possible chiral operators of the model are $u, \lambda_\alpha, S$ and $uS$. 
Here we mainly use symmetry argument, and it is possible that some of these operators disappear in the chiral ring (but
the $u$ and $S$ do not disappear because they can trigger nontrivial RG flows to free theories).
It would be very desirable that one can verify our proposal using
other methods. 

A 4d SCFT with finitely many chiral primaries seems to be rare among the theories known in the literature.
For example, such a theory cannot have any moduli space of vacua at all.
In analogy with 2d minimal models, we might call 4d SCFTs as minimal models if the theory has only finitely many chiral primaries.

\subsection{Central charges}
The central charges of the model can be computed by using the results of the $\CN=2$ AD theory \cite{Shapere:2008zf}.
In $\CN=2$ SCFTs, the central charges $a_{\CN=2}$ and $c_{\CN=2}$ are related to the 't~Hooft anomalies as
\beq
&\tr R_{\U(1)}^3=\tr R_{\U(1)}=48(a_{\CN=2}-c_{\CN=2}), \\
&\tr R_{\U(1)} R_{\SU(2)}^2=8(2a_{\CN=2}-c_{\CN=2})
\eeq
On the other hand, in $\CN=1$ theories, $a$ and $c$ are given in terms of 't~Hooft anomalies of the R-symmetry as \cite{Anselmi:1997am}
\beq
a=\frac{1}{32} (9\tr R^3 -3 \tr R), ~~
c=\frac{1}{32} (9 \tr R^3 -5 \tr R).
\eeq
Therefore, if we know $a_{\CN=2}$ and $c_{\CN=2}$, we can easily compute the central charges of the IR fixed point by using 
the relation \eqref{eq:Rc}. Such a calculation was done in \cite{Bolognesi:2015wta} in the case of the superpotential deformation $W \sim u$, 
and it was confirmed that the central charges are those of a free chiral multiplet in that case.

The central charges in the $\CN=2$ AD theory are given by \cite{Shapere:2008zf}
\beq
a_{\CN=2}=\frac{43}{120} \simeq 0.358,~~c_{\CN=2}=\frac{11}{30} \simeq 0.367.
\eeq
By using this and \eqref{eq:Rc}, we get the $a$ and $c$ of our model,
\beq
a=\frac{263}{768} \simeq 0.342,~~c=\frac{271}{768} \simeq 0.353.\label{eq:central}
\eeq
Note that the $a$ satisfies the $a$-theorem $a<a_{\CN=2}$ \cite{Komargodski:2011vj}.
Note also that the central charges satisfy the bound $1/2< a/c<3/2$ \cite{Hofman:2008ar}.
These facts support our claim that the superpotential deformation \eqref{eq:deform} leads to a nontrivial SCFT in the IR limit without accidental symmetries.

The value of $c$ obtained in \eqref{eq:central} is somewhat larger than the current lower bound $c \gsim 0.1$ obtained in \cite{Poland:2015mta}.
A possible reason is that our model has more than one chiral operator while the bootstrap method uses only one chiral operator. 
However, the central charges of our model are smaller than those of conventional SUSY QCD fixed points. 
For example, the central charges of the $\SU(2)$ theory with $N_f=4$ flavors
is given by $(a,c)=(3/4, 17/16) \simeq (0.75, 1.06)$.

\section{Discussion}\label{sec:discussion}
We have discussed an interesting model which has many desirable properties to be a minimal model.
However, there is still a gap between our model and the bootstrap constraints
and it is very interesting to close the gap from bootstrap side and/or model search side.

There are at least two ways to search for 4d minimal models. Let us describe preliminary results, leaving more detailed study for a future work.
The first method is to engineer four dimensional $\mathcal{N}=1$ theory by putting 6d $\CN=(2,0)$ theory on a punctured Riemann surface $C$ with
line bundles $(L_1, L_2)$ such that $L_1\otimes L_2=T^*C$.
\cite{Bah:2012dg,Xie:2013gma,Yonekura:2013mya} (see also \cite{Bonelli:2013pva,Giacomelli:2014rna}). 
To get $\mathcal{N}=2$ AD theories \cite{Xie:2012hs}, one chooses a sphere with one irregular puncture or a sphere with one irregular and one regular puncture, and the 
choice of bundle is $(L_1, L_2)=({\cal O}(-2), {\cal O})$.  To get $\mathcal{N}=1$ AD theory, one has more choices of bundles and punctures.

Let  us consider the case of 6d $A_1$ theory with $(L_1, L_2)=({\cal O}(-1), {\cal O}(-1))$. We use an irregular puncture of the following form
\begin{equation}
\Phi_1\sim {{ 1} \over z^{1+1/2}} \diag(1,-1),~~\Phi_2\sim {{1 } \over z^{1+1/2}} \diag(1,-1).
\end{equation}
The spectral curve of this system can be easily computed \cite{Xie:2013rsa}. By putting the puncture at $z=\infty$, we get
\begin{align}
& v^2=z+u_1,~ vw=z+u_2,~w^2=z+u_3.
\end{align}
We may use the 
automorphisms of the Riemann sphere to set $u_3=0$. We also
need to impose the relation $(z+u_2)^2=(z+u_1)z$, and this gives the chiral ring relation
\begin{equation}
u_1=2u_2,~~u_2^2=0.
\end{equation}
So there is a chiral operator $u_2$ satisfying the condition $u_2^2=0$. 
To find the scaling dimension of the operator $u_2$, we impose the condition that the canonical three form $\Omega=dv\wedge dw \wedge dz$ has 
scaling dimension 3. We find $\Delta(v)=\Delta(w)=3/4,\Delta(z)=3/2$, and $\Delta(u_2)=3/2$. 
These features are the same as the field theory model, and it will be very interesting to determine whether these models are the same or not.

Another direction is to study deformations of more general AD theories by a linear deformation $W \sim u$.
In this case it is not clear whether there is no accidental symmetry in the IR (see \cite{Bolognesi:2015wta} for some examples which flow to free theories). 
Assuming the absence of accidental symmetry, we find a model with smaller central charges. We start with $A_4$ AD theory, and turn on
 $W\sim u$ deformation with $\Delta_{\CN=2} (u)={10\over 7}$. We get $(a,c)=(633/2000,683/2000) \simeq (0.316,0.341)$~\cite{Giacomelli:2014rna}.  
It is very interesting to determine whether this theory really flows to a nontrivial SCFT without accidental symmetry, and if so, determine its chiral ring.

{\it Note added;} After the first version of this paper was submitted, the paper \cite{Buican:2016hnq} appeared which has overlap.
There, the relation $u \lambda_\alpha=0$ was also discussed.

\section{Summary}

Searching for minimal nontrivial quantum field theories is extremely interesting. In many cases the theories are intrinsically strongly coupled,
and conventional textbook approaches do not work at all, .
In two-dimensions, a beautiful story about minimal models has been developed starting from \cite{Belavin:1984vu} by conformal bootstrap methods,
and the studies in higher dimensions have begun recently.

In this letter we have initiated the search of minimal models in four-dimensional $\CN=1$ supersymmetric theories
by explicitly constructing a candidate.
We gave a possible definition of minimal models in terms of the finiteness of a certain class of operators (i.e., chiral primaries), and 
discussed some evidence that our theory satisfy that minimality condition. We have not given a complete proof of the minimality, but
a proof might be possible by using techniques such as in \cite{Buican:2016hnq}.
There is still a gap between our model and the bounds obtained by numerical conformal bootstrap~\cite{Poland:2011ey,Bobev:2015jxa,Poland:2015mta}.
The existence of such a gap suggests that there is room to learn something new about the nature of strongly coupled field theories in general.
Therefore, it is very important to fill this gap in the future.

\section*{Acknowledgments}
We would like to thank Dongmin Gang, Yu Nakayama and Tomoki Ohtsuki for stimulating discussions.
The work of K.Y is supported by World Premier International Research Center Initiative
(WPI Initiative), MEXT, Japan. D.Xie would like to thank the hospitality of IPMU where this 
work is initiated. The work of D.Xie  is supported by Center for Mathematical Sciences and 
Applications at Harvard University. 
%

\bibliography{ref}


\end{document}